\documentclass[12pt]{ejtex}

\usepackage{amsmath}
\EJissue{5}
\EJdate{Summer 2013}

\begin{document}

%Article environment
%usage: \begin{EJarticle}[optionalShortTitle]{title}{Author(s) - eventually // separated}{affiliations}{email}

%This article is for the Summer 2012 edition
%version 1.0 11/07/2012

\begin{EJarticle}[Quantum will]{Quantum Will:  Determinism meets\\ Quantum Mechanics}{Jos\'e I. Latorre}{Dept. ECM, Universitat de Barcelona, Spain\\
CQT, National University of Singapore\\ latorre@ecm.ub.edu}{Universitat de Barcelona, Spain} 

\begin{EJabstract}
We present a brief non-technical introduction to the standing discussion on the
relation between Quantum Mechanics and Determinism. Quantum Mechanics 
inherent randomness in the measurement process is sometimes presented as 
a door to explain free will. We argue against this interpretation. 
The possibility that Quantum Mechanics provides just an effective description 
of Nature which is only valid at our low-energy scales is also discussed. 
\end{EJabstract}

\section{The many faces of determinism}

Was it written at the time of the Big Bang that you would now be reading these words?
Is it already completely fixed the way we shall work, love and die? These questions
exemplify the emotional content unavoidably attached to any profound reflection on
determinism.\\

We may envisage a grand picture of the universe where physical laws act in a 
completely determined way, hence leaving no space for neither randomness, nor free will. 
We would then not be responsible for our wrong actions, neither for our successes, as both were
set to occur before our own existence. It would make no sense to doubt about a subtle
decision, given that our final choice was already fixed beforehand by the laws of Physics.
At stake are the rightness of justice, the root of morality and the essence of religions. It
is also under scrutiny our detailed understanding of Science and the laws that govern the universe.
No question that determinism is a subtle and controversial subject, constantly revisited,
that should be approached with care, respect and lack of any prejudice. What makes the
debate on determinism even more appealing nowadays is the constant and unstoppable
progress of Science. Biology, Chemistry and Physics are bringing new elements into the
discussion, hereby forcing a new, clean start on the analysis of determinism and free will.\\

The discussion on determinism in the form of causality is present in the history of
humanity from its very early stages. It is arguable that all ancient cultures looked for a
sequence of causal implications that end up with our reality. This chain of causes and
effects needed an origin or a cyclic universe, depending on the culture. Both solutions to
the initial problem of causes were often presented in poetic and allegoric terms, accepting
in a veiled way their unintelligibility. Yet, causality was accepted at intermediate steps. A
chain of causes and effects, with the possible inclusion of gods, described why we are hit
by a storm, why a drought damages the harvest or why sins will be inexorably punished.
Consequences always followed causes, and fate was
inexorable and
leaved no space for randomness and will. \\

Popular causality departs from scientific determinism in the sense that some amount of
unknown elements are accepted into the game. This subtle play between 
the ideas of an unavoidable future and some sparse disruptive non-deterministic
elements shows at the popular level the very same discussion which is at the heart of
scientific progress. Determinism seems real on every experiment we perform on
small controlled systems, but humans
remain reluctant to accept it at a large scale.\\

For determinism is deeply related to lack of free will. It follows from the strict
acceptance of physical determinism that there is no human freedom to alter the course
of events. The idea that we make a decision is just an illusion experienced by our brain.
The universe proceeds as an extremely complex machine, that has no chance of choosing
between options. \\

It is necessary to realize that the word ``determinism" 
has different meanings depending on authors and context. 
Often, determinism is qualified as ``causal determinism", 
``theological determinism", ``biological determinism", etc.
or may turned into lines of thought such as ``compatibilism".
Furthermore, determinism is often confused with predictability and the lack of determinism 
is associated to free will. It is fair to say that
determinism offers a rich landscape for theories, beliefs, arguments and refutations
for each of them. \\

It is thus safe to limit the present discussion
to debate the implications of Quantum
Mechanics on determinism, from a scientific point of view.
Many readers may prefer a more philosophical, ethical or even evolutionary approach
to the subject of determinism and should resort to the vast existing literature on the
subject~\cite{LatorreRef11,LatorreRef0}. On the other hand, some readers may be interested
in some of the conceptual developments which are taking place in the real of Quantum 
Mechanics, in particular when a consistent and critical attitude is taken.

\section{Complexity remains deterministic}

Complexity conditions our relation to physical phenomena. We experience on a daily
basis the apparent impossibility of predicting the outcome of our actions. For instance,
we may stir the cream topping a cup of coffee and marvel at the changing black and white
figures we have created. It looks out of question that we could predict such a detailed
phenomena. Let alone the problem of weather forecast or the crashing of a tall wave on
a beach. The number of particles involved in these phenomena is enormous, so is the
number of coupled equations to describe them. Complex phenomena look unpredictable
to the human mind. \\

On the other hand, it should be clear that complexity is not related to lack of determinism. 
The absence of sufficient computational power to describe complex phenomena
limits our capacity to predict in full detail very many macroscopic phenomena. Yet, this
is a practical, not an essential obstruction to predictability. It is also an anthropomorphic
way of reasoning, since we confuse determinism with human predictability. The Earth
will follow a very precise path around the Sun, whether humans compute its trajectory or
not. Furthermore, there is no doubt that our predictions are improving as computers get
faster since the Newton laws that describe the motion of particles are well understood.
It is actually a wonderful fact that the detailed understanding of Newton’s deterministic
laws opens the possibility for engineering, that is, for the instrumental use of such laws
by humans. It is thus possible for us to tame solids and make bridges, cars or clocks.
We can build skyscrapers, we can correct our vision with glasses, we can travel to the
Moon. We understand the laws and exploit them. \\

A separate but related problem to computing the evolution of a complex system is
that the precise knowledge of the correct differential equations that control a system is not
enough to fully describe its evolution. Indeed, to predict how a system evolves it is also
necessary to have a perfect knowledge of initial conditions. Any mistake in introducing
the initial conditions of all the particles of a system may completely spoil our prediction.
As a matter of fact, it is well-known that non-linear differential equations exhibit chaotic
behavior. That is, small departures of a given initial condition grow into exponential
differences as time evolves. No matter how precise we try to get the initial conditions, we
shall always have a finite accuracy and the evolution of the system will be unpredictable
at large time scales. Therefore, the evolution of chaotic systems can only be predicted
in practice on a short time basis. Weather forecast is a chaotic problem. This is why it
is not possible to accurately predict whether it will rain in London in a month time. We
may say that chaos means that some systems need much more computational power and
observational effort to be predicted than others. But, the profound issue of determinism
remains unaltered. Equations are valid, initial conditions are there (whether we know
them or not) and the future is completely fixed, even though humans need enormous
computational efforts to read it.\\

The previous discussion about the essence of determinism, as oppose to the apparent
stochasticity of our environment was at the heart of the original discussion by 
Laplace~\cite{LatorreRef14}. 
A thoughtful approach to determinism can be traced back to ancient Greece, as
well as other ancient cultures. But it is Laplace the first who noted down the inexorable
dictate of differential equations on the flow of events. His poetic words described past
and future as present for a powerful mind that would have infinite computational power.
He also argued that our understanding of the motion of stars provides a glimpse for such
an idea.\\

What Laplace understood is often hard to get for many people. The essence of
determinism is not about complexity, lack of knowledge of initial conditions, or even
the poor understanding of laws of Nature. Determinism means that there are laws that
will exactly control the evolution of the universe, to every minute detail. It is irrelevant
whether we can compute them or not, whether we can measure the initial conditions,
even whether we know the laws at all. Humans play no role.\\

\section{Biology and free will: the meaning of justice}

The above discussion has a profound consequence on free will. Taken at face value, strict
determinism of the laws of Nature leaves no space to free will. This is indeed shocking.
As Einstein, we are ready to think that the moon, quite a macroscopic object, is not
deciding whether to rotate around the earth of not. But it is hard to accept that neither
humans have any options.\\

Recent progress in Molecular Biology is producing a large impact 
on our understanding of decision making. 
Some experiments show that decisions can be anticipated several seconds ahead
by monitoring the chemical activity of the brain.
It is then possible to envisage apparent free will as complex
workings of our brains. Growing evidence shows that decisions are reduced to flows of ions, 
whose origin are yet to be fully understood, but might  
lie on the realm of Chemistry!~\cite{LatorreRef15, LatorreRef16, LatorreRef19, LatorreRef9}.\\

This discussion and its consequences have been put forward by Cashmore~\cite{LatorreRef7}. The
understanding of biological events is currently based on three conditioning elements: 
Genetics, Environment and Stochasticity (often referred as GES). Note that neither Genetics
nor Environment do have any bearing on the core issue of determinism. Both Genetics and 
Environment may bring complexity, but they follow strict causal laws. On the other hand,
Stochasticity, if understood as an inherent element of randomness in Nature, would completely
spoil determinism. Before analyzing Stochasticity, let us reflect on the moral debate that
follows the possible reduction of free will to pure Chemistry.\\

Let us accept that decisions are just brain processes which can be monitored using
fMRI and, furthermore, can be anticipated given our understand of chemical reactions.
This implies that humans have no choices. Each decision is the result of deterministic laws, 
and the outcome can be anticipated provided we monitor the chemical reactions in
our brain. Some subjects will behave in some way,
some in others, but the differences can be predicted in advance. We may think of  decision making
as a magnifying process, whereby small chemical differences get enlarged to macroscopic
distinct events. Our behavior, whatever complex it seems, does not involve any
genuine randomness at molecular level.\\

This is indeed a unsettling possibility.
Whether a person misbehaves or acts according to accepted moral rules would be pre-fixed~\cite{LatorreRef7}. 
Then, we may ask what is the meaning of punishment? Why do we have prisons? Should justice consider
preemptive actions against individuals who are predicted to misbehave? These moral
issues have devastating effects on the organization of our society and need profound
reflection. We may argue in a pragmatical way that our laws are there to make safer the life of some, 
given the preconditioned behavior of others. Yet, even this process of creating laws and imposing 
them would be part of the deterministic global evolution of the universe. Whatever scape we may try, 
determinism will be there to reassess that the laws of Nature are complex 
enough to make us feel a false sense of freedom, where there is none.\\

Free will may well be an illusion, but it may be a very useful one from 
an evolutionary point of view! This departure of the main line of discussion of free will in Neuroscience and
Physics is intriguing and would need a separate discussion for itself~\cite{LatorreRef6}.

\section{Randomness does not imply will}

Determinism seems to be eliminated altogether if some randomness 
is present in the laws of Nature. Indeed, if at any intermediate stage of a 
physical process there is a random event, there will no way to determine a priori the
outcome of such a process. \\

Randomness is then often quoted as an ingredient for free will. It should be 
noted that this is a very weak reasoning. Randomness lacks, by definition, any sense of will.
There is no foreseeing of the future, neither intention associated to the drawing
of a dice. In contrast, free will implies evaluation of options and foreseeing of 
the possible outcomes associated to each available choice. Free will is about consciousness,
moral stands, and drive. Quite on the contrary, 
 randomness has no will. We shall come back to this point within the realm
of Quantum Mechanics later on.\\

At this level of the discussion it is essential to analyze very critically randomness.
There are models in statistical physics that incorporate elements of randomness. But,
we should be aware that these models are effective theories that were constructed to describe averages.
A trivial example to exemplify this point is gambling at a roulette.
We are told that a roulette has equal chances to have all possible outcomes. 
But it is clear that the final output is a consequence of 
the speed of the ball, 
trajectory and details of the machinery. It is so clear to the owners of the casino
that they perform very strict tests on each roulette before it is used for betting.
It is implicitly understood that a very detailed analysis of constituents would supersede
any random description of the roulette, though the computational cost of such a step would be gigantic.
As a consequence, we should be aware that not all randomness is such. We use
the word random to mean that the computation of an outcome is out of our present capabilities. \\

Quantum Mechanics will introduce an enormous twist in this discussion.

\section{Determinism goes elementary}

The inevitability imposed by determinism depends on the crucial issue of having
no randomness associated to the laws of Nature. This is the stochasticity part of the
GES discussion. If the basic laws of Nature do present some intrinsic randomness, then,
determinism is no longer ruling the world. This question is outside the realm of Biology
and of Chemistry which do not deal with the laws of Nature at
the most fundamental level. We need to go deeper into the key axioms of Physics.\\

At present we have discovered four apparently different types of interactions that
control the behavior of the most elementary particles which are observed in our most powerful
machines, the elementary particle colliders.
These four interactions conform the so-called Standard Model and correspond to Strong
interactions (described by Quantum Chromodynamics), Weak and 
Electromagnetic interactions (described by the Quantum Electroweak theory) and Gravitational interactions
(where the best we can do nowadays is to use Einstein’s General Relativity). No present
experiment seems to suggest the existence of new laws, neither the failure 
of the Standard Model. There are, though, indications hinting at possible unification schemes of
the known interactions at higher energy regimes. For instance, the Standard Model does
not predict the masses of each elementary particle, neither why quarks have charges with
exact fractions of the electron one. It is likely that we have a partial understanding of
the laws of Nature, adequate to the scales we are able to probe. At higher energies, the
laws of Physics as we understand them may merge or be substituted with a more fundamental ones.\\

It is a remarkable fact that all interactions but gravity have been understood as 
manifestations of the basic scheme provided by Quantum Mechanics. The axioms of
Quantum Mechanics, namely ideas such as quantum superpositions (that follow from
the Hilbert space structure for quantum states) or projection of the wave function (as
a projection of the information we have on a system when a measurement is done) are
routinely checked when predicting the collisions of elementary particles at large accelerators. 
The same quantum laws are checked daily on every experiment in many branches of
Physics: nuclear physics, condensed matter physics, astrophysics, quantum optics, etc.
So far, every single experiment is consistent with Quantum Mechanics. Indeed, experiments
make constant progress on the understanding of the details of the forces, but do exhibit
perfect agreement with the underlying structure provided by Quantum Mechanics.\\

It is important to emphasize that we understand the basic laws of Nature with great
detail and accuracy. For instance, Quantum Electrodynamics -a piece of the Electroweak
theory- sustains at large our amazing technological progress. We can master electrons to
travel through circuits, jumping over transistors, to create computers, phones, or any of
the many devices that serve to ease our life. We also master photons to travel through
optical fibers, coding information. Amazingly, we also master the interaction of single
electrons with photons at the level of producing quantum logical gates.
All of these technological marvels are originated from
our understanding of the interaction between matter and light. Recent progress includes
monitoring of entanglement to obtained clock with a precision of one second in the age
of the universe. There seems to be no limit in the ways we shall exploit Quantum
Mechanics.

\section{Effective theories}
Along the history of Science, many theories have superseded previous ones. Theories that
were considered final were shown to only describe a part of the phenomena they were
supposed to govern. A better theory would then emerge, a theory to describe a larger
range of phenomena with better accuracy and, often, in a simpler way. As an example,
we may recall the Ptolemaic theory describing the motion of planets. It is said that
Alfonso X the Wise, King of Castile, responsible of the Alfonsine tables and founder of
the Toledo School of Translators, was taught Ptolemaic theory. At his time, the motion
of a few bodies was described using over 70 cycles and epicycles. The king’s probably
apocryphal reaction was to say that, if he were God, the theory would be simpler. As
a matter of fact, Copernicus made it simpler by understanding that the natural center
to view the orbits of all the planets was the Sun rather than the Earth. Later, Newton
created a theory of gravity that would describe with a single equation the falling of bodies
and the motion of planets. The theory was more precise, described apparently unrelated
phenomena with a same law, and was aesthetically beautiful. The latter twist on gravity
corresponds to Einstein's theory of General Relativity. There, a higher symmetry principle
describes gravitational phenomena in a range of energies larger than Newtonian laws.
We know that Einstein's General Relativity is not the ultimate formulation of gravity,
since it does not accommodate quantum phenomena. String Theory is the only serious
but incomplete attempt for Quantum Mechanics to encompass gravity.\\

The lesson to be learned from the improvement of our description of gravitational
phenomena is that our knowledge should always be considered as partial and open to
improvement. Our theories are just effective theories. That is, we use effective laws
that describe all known phenomena within the range of experiments which are available.
In the case that new phenomena prove these laws to be of limited validity, scientists
would then look for an extension or recreation of a new theory, that should in turn be
considered as an effective one.\\

The summary of the above discussion is simple: as far we can tell Quantum Mechanics
rules Nature, though we should keep our mind open to some better theory. So, there are,
at least, three reasonable questions to be asked on the relation of Quantum Mechanics
to determinism. First, is Quantum Mechanics deterministic? Second, can Quantum
Mechanics explain free will? And third, is it possible that Quantum Mechanics may
emerge from a more fundamental and deterministic theory? We shall discuss these
questions along the next sections.

\section{Probabilist results of measurements in Quantum Mechanics}

Quantum Mechanics establishes that the outcome of a measurement is 
genuinely probabilistic. Therefore, Quantum Mechanics seems to rule out determinism. To be precise,
the postulates of Quantum Mechanics state that observables are related to operators in
the theory, that the result of the measurement must be one of the eigenvalues of the
operator, and that the actual result which is observed in an experiment will occur with
a probability which is computable from the wave function that describes the system.
We refer to the quantum postulate of how random outputs are computed from the
state of the system as Born rules.
This may sound too technical for non-experts, but the key element is that results of
measurements are inherently random. To be absolutely clear, the only element
which is not fully predetermined takes place at the act of performing a measurement.\\

It is fair to mention that the present axioms of Quantum Mechanics are not 
uniquely formulated. Actually, there are alternative ways to set the foundations of
Quantum Mechanics  which are perfectly consistent~\cite{LatorreRef10}. 
On the other hand, it is unlikely that any of
the reformulations of Quantum Mechanics that we may envisage will ever get rid of its inherent 
randomness. We may argue that randomness is conserved: either
it is there or not. Therefore, the relation of determinism to Quantum Mechanics is
at a deeper scale than the discussion made in a particular set of axioms, for the
relevant fact is that Quantum Mechanics brings the ingredient of inherent randomness. \\

Let us emphasize again that Quantum Mechanics only brings an element on randomness
when measurements are performed. Between measurements, the evolution of a quantum
system is described by the Schr\"odinger equation, which is a deterministic differential
equation for the wave function that contains the information on the system. Furthermore,
measurements need an observer, that is, some part of the system is taken apart and acts
as observer of the other part. This is certainly an uncomfortable setting, which has been
extensively discussed. On one hand, Quantum Mechanics produces fantastic predictions
but, on the other hand, our understanding of the role of the observer is distressing.\\

It is not easy for a scientist to accept randomness in the measurement process. 
Einstein's dislike for this quantum postulate is well-known. He would argue that God does
not play dice in the atomic world (see for instance the wonderful exchange of letters
between Einstein and Born~\cite{LatorreRef4}). Einstein would not accept the 
need to resort to a random
number generator to describe Nature. Instead, he defended a local element of objective
realism which prefixes the results of experiments. The discussion initiated by Einstein is
a wonderful history of scientific honesty. It was much later when Bell sorted out the
right way to proceed~\cite{LatorreRef3}. Bell realized that local realism can be experimentally verified or
disproved. The idea is that local realism can not accommodate correlations 
among observations at separate sites which that are present in the real world.
He proposed the so-called Bell inequalities that
would tell apart whether our world follows the rules of local realism 
or those of Quantum Mechanics. 
Following Popper, Bell showed that classical physics can be falsified 
in favor of Quantum Mechanics.
The first experiments by Aspect~\cite{LatorreRef2} showed that our world is quantum,
whether we like it or not. At present, Bell inequalities are checked by freshmen students in Physics schools.\\

There is a recent relevant contribution to randomness in Quantum Mechanics 
{\sl vs.} randomness in Classical Physics. The basic idea goes as follows. 
The question is whether it is possible to distinguish
a truly random series from a deterministic series that imitates the former one.
For instance, we may open a Casino
and buy a random number generator to run a slot machine. Can we be certain that this machine
produces a truly random series of numbers? As a matter of fact we cannot. It may
happen that the machine we bought has some inner deterministic generators of numbers
that simulate a random series. We could check the machine for a few millions outputs
and verify that the distributions obey the expected rules of a random generator. But,
then, on the opening of the Casino, the next output in the slot will be perfectly determined by
the person who wrote the deterministic algorithm. The new development
in this discussion is that the lack of distinguishability between true and fake randomness
no longer holds in Quantum Mechanics. We can use entangled states and Bell inequalities to
generate truly random numbers. This subject is subtle and is fully explained in~\cite{LatorreRef17}.
There is yet an unsettling premise. Full quantum randomness 
comes at the price that experimenters must be free to choose instantly the 
measurement they perform. This initial randomness in the choice of measurements
might be very small but not zero for quantum randomness to emerge full glory.
We shall now discuss this fact in some more detail.

\section{Random choice of experimental settings}

Bell inequalities are at the heart of the randomness discussion as we have been arguing
previously. Let us briefly recall how an experiment testing a Bell inequality
 is organized. The typical set up for this verification of Quantum Mechanics
requires a source of pairs of particles, and each one of them flights to a
different observer, usually named Alice and Bob. In order to analyze the properties of the state,
Alice and Bob separately proceed to measure some property of the particle they have
separately received. To be more concrete, let us consider the case where
two photons are created with correlated polarizations and sent one to Alice, an the
other to Bob. Then Alice and Bob may choose to measure the polarization of the
light using polarimeters pointing in different directions. The statistics obtained when
accumulating the results over very many random directions for the polarimeters turn
out to be consistent with Quantum Mechanics but not with classical physics. Indeed,
entangled quantum states are much more correlated that any classical state, as proven in
the laboratories. These experiments allow us to rule out theories where a local element of
realism describes deterministically the output of each experiment. Einstein was wrong,
but his contribution was essential to consider correlations as the figure of merit that
discriminates a quantum from a classical world.\\

Everything seems good and fine for Quantum Mechanics. Yet, there is a small but
deep cave at. All along the Bell experiment, both Alice and Bob must take the decision
to point their polarimeter in some direction at the very moment 
they perform their measurement. So we must accept some free will on the
side of Alice and Bob! If we could predetermine the polarimeters direction that Alice
and bob will use, then it is possible to create a local hidden variable model, consistent
with classical physics~\cite{LatorreRef13, LatorreRef12,LatorreRef32, LatorreRef33}.\\

The problem of freedom of choice for experimenters is once again showing our poor
understanding of the role of the observer in Quantum Mechanics. The observer is 
responsible for the directions of the polarimeters, for the collapse of the wave function
and for the probabilistic outcome of the experiment. In its absence everything would be
described as deterministic evolution of the wave function.

\section{No Quantum Will}

Let us accept for the time being that Quantum Mechanics does bring
unquestionable randomness. Then, we can argue that
Quantum Mechanics seems to kill determinism (with the cave at that we 
must attribute some freedom of choice
to observers). This is good news for the many people that hate the negation of free
will. Nevertheless, it is unclear in which sense the measurement quantum axiom favors
free will~\cite{LatorreRef5}. We may argue on the contrary in two different ways. 
First, we may consider
that all the physical interactions along a decision process that takes
place in our brain are described by a unitary and deterministic evolution, as dictated by
the Schr\"odinger equation. In a decision process, no measurement would ever be made.
Thus, Quantum Mechanical evolution of the whole system remains deterministic! A
second counterargument of profound consequences says that true randomness is not free
will. Outcomes may well follow random rules, so they are not determined, yet there is
no choice to be made by any part of the quantum system. 
Quantum randomness follows no will.\\

The issue becomes now really subtle, so let us slow down and reconsider the quantum
mechanical implications on determinism and free will. So far, we have enumerated a
number of key ideas related to the possible absence of determinism, namely  complexity
and randomness, and we look for some understanding on free will. We argued previously that complexity 
has no bearing on determinism. Free will does correspond to the moral 
human appreciation of the whole problem and, as such, is the ultimate big question. Yet,
Quantum Mechanics postulates do humbly discuss the role of randomness in the 
measurement process. Moreover, quantum randomness can only be separate from classical
randomness in Bell-type settings. But then, an initial element of randomness must be
granted to observers. Altogether, Quantum Mechanics is about randomness, not free will.
We argued above that randomness is postulated for the result of quantum experiments. 
This leads to a peculiar conclusion. If free will is eventually emerging from
Quantum Mechanics, we are forced to grant such a free will to a distribution of 
particles and measurements.\\

This latter idea is the basis for the so-called
 Free Will Theorem~\cite{LatorreRef8}, and its evolved
version Strong Free Will Theorem. This theorem provocatively uses the word free will
for the indeterminism inherent to local quantum degrees of freedom. Based on some
basic axioms (FIN, TWINS and MIN), it is proven that the outcome to local observables
is not dictated by previous properties of the system.
The discussion above shows clearly the delicate use of the word ``free will" in the context
of Quantum Mechanics. Any honest discussion about the absence of determinism in
Quantum Mechanics should also include the point that randomness does not explain free
will. A strong form for this comment would be to disregard as erroneous the concept of
Quantum Will.\\

\section{Quantum Mechanics as an effective theory}

Many people will argue that Quantum Mechanics has opened a window for free will, given the
non-deterministic postulate of measurement. Many, on the contrary, will consider that
Quantum Mechanics is reinforcing the absence of free will, since evolution is deterministic
and measurement is genuinely random. In the extreme case, it is possible to argue
that Quantum Mechanics should be applied on the universe as a whole and, then, no
observation would ever be made. What we view as an observation in our labs is nothing
but a deterministic evolution within the total system. At the grand scale we are finding a
familiar situation, the system may be too complex for humans to predict using Schr\"odinger
equation, but we are certain that a unitary an deterministic evolution is taking place.\\

A more profound and far-reaching discussion is to consider Quantum Mechanics as
an effective theory~\cite{LatorreRef1}. 
The idea was already discussed in the context of the Standard
Model but now takes a new twist. We may find that the four basic interactions are
just an effective way of describing Nature, and that a larger symmetry will encompass
these four theories. But we may go much further away. We may find that the quantum
structure underlying the Standard Model is effective! That would mean that the 
quantum axioms
that we use are just a good approximation to a deeper organization of concepts. This is
the approach to Quantum Mechanics as an effective theory suggested 
by  't Hooft~\cite{LatorreRef18}.\\

The theory proposed by 't Hooft is somewhat
reminiscent of the old discussion about initial conditions. A deterministic system
will be fully predictable given some perfectly known initial conditions. On a twist of
the above argument,  't Hooft has argued that the distribution
of classical initial conditions of a cellular automata which are consistent with a given
output is at the origin of using probabilities in a quantum mechanical way to describe
such an automata. That is, Quantum Mechanics probabilities may only reflect the sets
of classical initial conditions which are compatible with our experiments.\\

The argumentation by 't Hooft links the odd understanding we have of Quantum
Mechanics to the also odd understanding we have of Gravity. At present, we do not
know how Gravity is realized at very high energies, that is at the Planck scale 
($10^{19}$
GeV). At those energies, Gravity must merge with Quantum Mechanics and the output
is simply not known. It may happen that space and time are no longer differential
manifolds, it may be the case that space and time are emerging structures from 
more basic ingredients. We really
do not have the faintest clue. The proposal by 't Hooft argues that at the Planck scale, Nature
is controlled by a deterministic theory such as cellular automata, whose elements have
the Planck size. Apparent randomness is just an spurious effective description of
physical phenomena, only useful 
at our present low energies.

\section{Conclusion}

Some of the discussions we have presented can be summarized in short statements. Let us start
with clearing the path from complexity:
\begin{itemize}
\item Complexity is irrelevant for determinism.
\item Predictability or human knowledge of the laws of Nature is irrelevant for determinism.
\end{itemize} 
Determinism is related to the existence of exactly obeyed laws. Whether we know them
or we can compute them is not the point. \\

Then, a human question
arises: if Nature follows deterministic evolution, is there room for free will?
The answer should be ``no". It is the hands of Chemistry and Biology 
to clarify this fundamental issue 
and go ahead with systematic studies of decision making and consciousness at
molecular level. It may well be the case that 
free will reduces to an illusion. Decision making would be a magnification process
that could be
predicted on  a purely chemical basis, whatever complex this process may be.\\

Yet, are we sure that Nature follows deterministic laws?
We know a fact:
\begin{itemize}
\item The only possible source for indeterminism in our present understanding of the laws of Nature
is the measurement process in Quantum Mechanics.
\end{itemize} 
This measurement process is, 
sadly enough, the less understood piece of Quantum Mechanics. Whether quantum randomness 
(Born rules) and the collapse of the wave function will stay as part of the ultimate
understanding of Nature is far from obvious. In any case
\begin{itemize}
\item Quantum randomness follows no will.
\end{itemize} 
Will implies human concepts as consciousness which are not part of the Born rule.
Inherent quantum randomness rules out perfect predictability, but it gives no space for will.\\

The discussion of randomness in Quantum Mechanics is heavily affected by Bell inequalities
and non-locality of correlations. Those are very subtle issues that may be better
understood in the future.
Furthermore, we should always be aware that
\begin{itemize}
\item Quantum Mechanics may well be an effective theory.
\end{itemize} 
It is possible that Quantum Mechanics is no longer the right way to describe Nature
at very high scales. If so, some quantum features as randomness could be an emerging
effective description, a substitute for a deeper non-random description of Physics.
It may be argued that
\begin{itemize}
\item Quantum Mechanics will get severely modified when merging with Gravity.
\end{itemize} 
The structure of the would-be unifying theory of Quantum Gravity is simply unknown. 
We may argue that the ultimate theory will be purely deterministic 
-a theory of cellular automata- but we may also
argue in the opposite direction.
It is fair to 
accept that the hope that the main problems in Physics, namely the emergence of
quantum randomness and the quantization of gravity, will be solved simultaneously
is probably naive.\\

We may also wonder what are the coming developments in relation to determinism
(it does sound funny to determine the next steps to understand determinism!). It is likely
that the relevant near future investigation in the ever lasting discussion 
on free will and determinism will
be centered in the understanding of decision making at the chemical level. The systematic
study of brain processes will eventually clarify to what extend will is predictable and
whether consciousness is also reducible to complex chemical reactions. There is no doubt
that progress in the field of Neuroscience will produce a profound reassessing of some
controversial intellectual positions about determinism. In a way, we may think of the
impact of fMRI experiments on free will as the equivalent of Bell inequalities on local
realism. We may have strong opinions, but experiments will decide.\\

As a separate line of research, it is clear that we need proposing and analyzing theories
that would produce Quantum Mechanics as an effective description of our world. This
may be considered as too far fetched, since such theories are not likely to be falsifiable. Yet,
mathematical consistency has proven an extremely powerful tool to discover the rules of
Nature. More realistic progress should come from analyzing in depth the emergence
of probabilistic rules in Quantum Mechanics, a subject that relates to decoherence
and quantum darwinism~\cite{LatorreRef31}.\\

On a personal final note, let me add that I join the group of scientists
that argue that the ultimate description of Nature must arise from
Arithmetics. But this is an even more profound subject that touches 
aesthetics and -if possible- absolute truth, for which we have no clues.

\end{EJarticle}

\end{document}